# Lasing Mode Control in ZnO Nanowire Lasers Coupled to $TiO_2$ Nanopillars

**Daniel Repp[1,*], Francesco Vitale[2], Raja Hoffmann[2], Isabelle Staude[1,2], Thomas Siefke[1,3], Carsten Ronning[2], and Thomas Pertsch[1,3]**

[1]*Institute of Applied Physics, Abbe Center of Photonics, Friedrich Schiller University Jena, Albert-Einstein-Straße 15, 07745 Jena, Germany*
[2]*Institute of Solid State Physics, Friedrich Schiller University Jena, Max-Wien-Platz 1, 07743 Jena, Germany*
[3]*Fraunhofer Institute for Applied Optics and Precision Engineering, Albert Einstein Straße 7, 07745 Jena, Germany.*
*daniel.repp@uni-jena.de

**Nanowire lasers based on spectrally broad gain materials typically exhibit multimode emission, which limits their utility in applications requiring precisely defined spectral characteristics. In this study, we experimentally demonstrate the selective, narrow-band suppression of individual lasing modes in zinc oxide (ZnO) nanowire lasers through near-field coupling to titania ($TiO_2$) nanopillars. Our study is based on photoluminescence spectroscopy in the lasing regime, which is complemented by finite-difference time-domain simulations. We show suppression of longitudinal nanowire lasing modes by their resonant coupling to the whispering gallery modes in nanopillars of suitable size. In addition to spectral selectivity of the coupling, we reveal two distinct coupling regimes based on polarization matching between whispering gallery modes and nanowire waveguide modes. These findings demonstrate a promising approach towards engineering the spectral emission properties of nanowire lasers through tailored photonic architectures.**

Zinc oxide (ZnO) nanowires can function as nanoscale lasers, combining optical gain, waveguiding, and resonator feedback within a single monolithic structure [1–3]. Their diameters can be reduced to sub-wavelength dimensions [4], facilitating near-field coupling to photonic structures and enabling control over key emission characteristics [5–9]. However, the integration of nanowire lasers into high-precision photonic systems is often hindered by the occurrence of multimode lasing under continuous-wave (CW) or quasi-CW excitation, typically resulting in the simultaneous excitation of several longitudinal waveguide modes [10–11]. Previous studies have demonstrated that nanoparticles in the near field of nanowires can influence their waveguiding behavior [12-13]. In this letter, we experimentally investigate a coupled nanowire–nanopillar system by micro-photoluminescence (μ-PL) spectroscopy in the lasing regime and demonstrate that dielectric $TiO_2$ nanopillars can induce selective suppression of individual longitudinal modes in ZnO nanowire lasers. To support our experimental findings, we perform comprehensive finite-difference time-domain (FDTD) simulations, quantifying waveguiding losses in the nanowire–nanopillar system and distinguishing between absorptive losses in the nanopillar and additional scattering contributions. We find that the observed suppression effect is primarily attributed to resonant, polarization-sensitive coupling between nanowire lasing modes and spectrally narrow whispering gallery modes of the nanopillars, signified by increased overall absorption as the $TiO_2$ bandgap closely overlaps with the ZnO emission range [14–15]. Depending on the mode excited in the nanopillar, scattering is either enhanced or reduced.

The samples investigated in this article were fabricated by ion beam sputter deposition (Ionfab 300LC by OIPT) of a layerstack consisting of 70 nm Al, 20 nm $Al_2O_3$, 200 nm nanocrystalline $TiO_2$ and 80 nm chromium on a silicon substrate (100 mm, SSP by Si-Mat). The aluminum layer was deposited to investigate possible effects of the nanowire/nanopillar system on the coupling between nanowire lasers and surface plasmon polaritons [16] in future work. Afterwards, the titanium oxide layer is patterned into circular nanopillars by character projection electron beam lithography and inductive coupled plasma etching (SI500C by Sentech Instruments GmbH) using the chromium layer as an intermediate hard mask. Finally, the chromium was removed by wet etching.

Two samples with different layouts of the nanopillars were fabricated, denominated as Layouts A and B. Layout A has a grid of 4x4 groups of nanopillars, each group consisting of 20x20 nanopillars spaced by 50 μm in all directions in a rectangular grid. In every group, the nanowire diameters are identical up to fabrication tolerances and vary between groups. Layout B has a grid of 4x4 groups, where every group is identical, consisting of 16x16 nanopillars. For Layout B, the nanopillars within one row have identical diameters, but different diameters for every row. For Layout B, rows are staggered by 25 μm in both directions to minimize nanowire movement between nanopillars when performing nanomanipulation. A sketch of the layouts is shown in Fig. 1. While Layout A makes it possible to reliably characterize the effect of specific nanopillar diameters to find nanopillar sizes suitable for coupling to nanowire lasers, Layout B offers the possibility to quantitatively explore these coupling effects by targeted moving of a single nanowire between differently sized nanopillars.

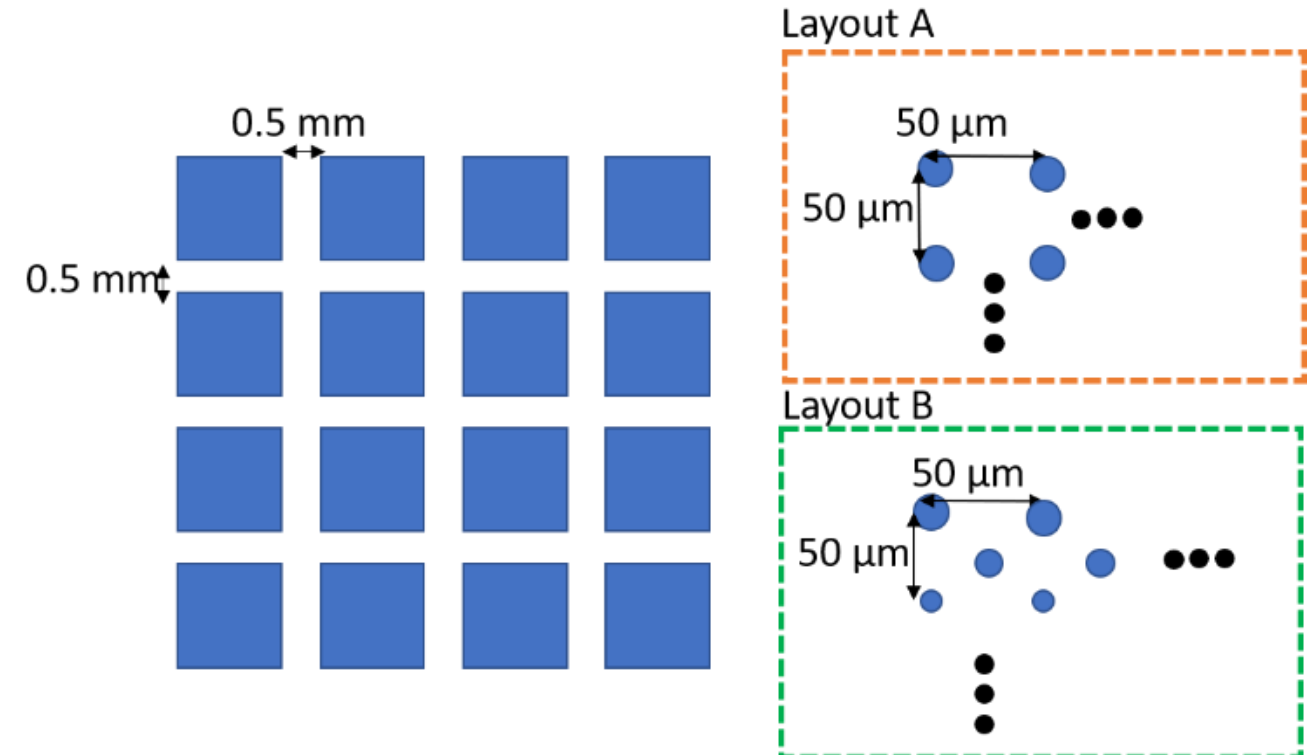

**Fig. 1** Sketch of the chip layouts. All chips are covered by a 4x4 grid of nanopillar groups. For Layout A, every group has 20x20 nanopillars of identical diameters with different diameters for every group, whereas Layout B has identical groups consisting of 16x16 nanopillars with rows of identical diameter and a change of diameter from row to row.

In Fig. 2a an SEM image of one nanopillar is shown before the removal of chromium and Fig. 2b shows a focused ion-beam cross-cut of the structure. More details about the fabrication process can be found in Supplement 1, Sec. 1.

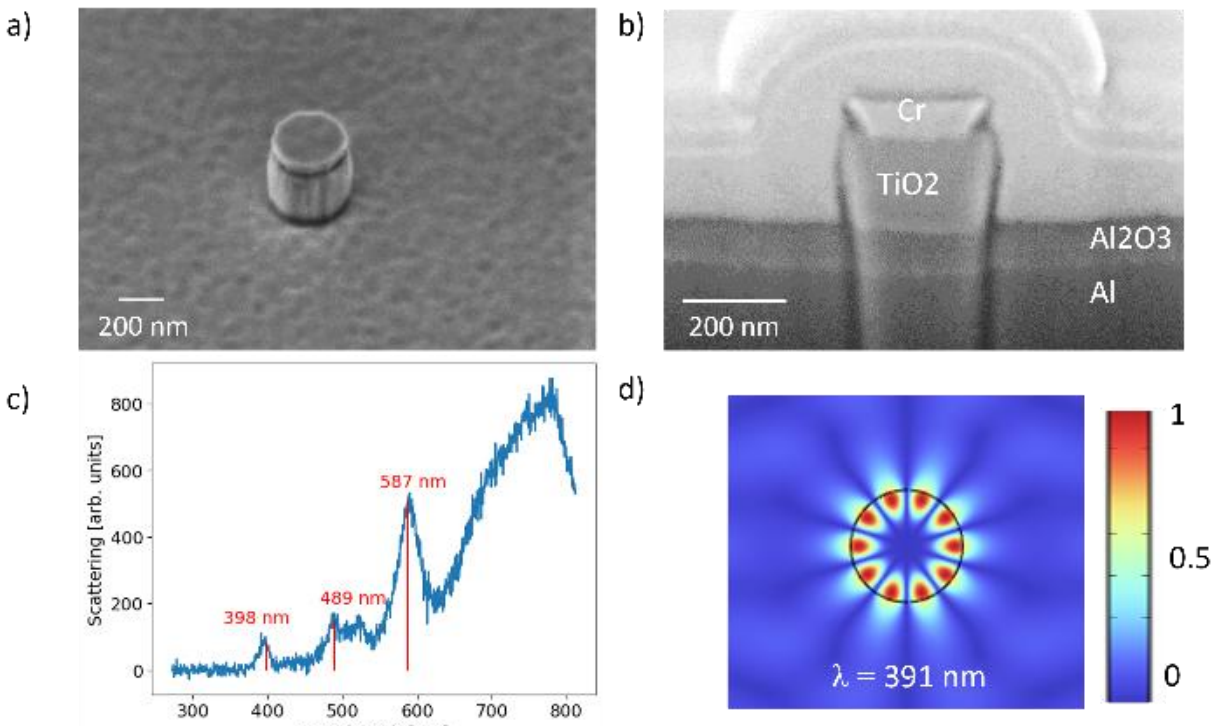

**Fig. 2** a) SEM image of a titania nanopillar (before removal of the chromium hardmask to provide better conductivity for the SEM images). b) FIB cross-cut of a titania nanopillar. c) Scattering spectrum of a single nanopillar with a diameter of 340 nm measured as the difference between light scattered by the nanopillar and the substrate, with red lines indicating the spectral position of the modes as calculated by Finite Element Method. d) A electromagnetic mode of a titania nanopillar with a diameter of 340 nm close to the observed scattering maxima calculated by Finite Element Method.

The nanopillars were characterized using a custom-built microscope setup. A halogen lamp was used to illuminate the sample in dark-field mode. The scattered light was then collected through the microscope and an adjustable knife-edge aperture was used to isolate and image the scattered light from a single nanopillar. Then, the light was focused through a microscope objective into an optical fiber and the spectrum was recorded with a Horiba 320i spectrometer with a grating of 300 l/mm and a blaze angle optimized for a wavelength of 500 nm. The measurement was repeated for a part of the sample containing no nanopillars and the resulting spectra were subtracted, as shown in Fig. 2c. For a nanopillar with a diameter of 340 nm, scattering resonances were observed at wavelengths of 398 nm, 489 nm and 587 nm. The spatial distributions of the electric field magnitude of the optical modes corresponding to the scattering resonances closest to the nanowire emission wavelength of 385 nm are calculated by Finite Element Method (COMSOL). A whispering gallery mode is found and shown in Fig. 2d, revealing that close to the ZnO emission frequency around 385 nm, whispering gallery modes can be supported.

To investigate the interaction between individual ZnO nanowires and $TiO_2$ nanopillars, we performed single-nanowire photoluminescence (PL) measurements combined with precise nanomanipulation of the nanowires' position with respect to the nanopillars. The integration of a nanomanipulator into the micro-PL (μ-PL) setup allowed for controlled repositioning of individual nanowires, which had been pre-transferred onto the samples using a dry imprint technique. This setup enabled us to bring a single nanowire into contact with a specific $TiO_2$ nanopillar and subsequently retract it, allowing for a systematic study of the influence of the nanopillar on the nanowire lasing emission. Power-dependent PL spectra were recorded at room temperature for each configuration, providing direct insight into the optical response of the coupled and decoupled nanowire-nanopillar system. More details about the experimental setup can be found in Supplement 1, Sec. 2.

First, the PL emission of a ZnO nanowire with a diameter of 255 nm originally imprinted several micrometers away from any nanopillars on a sample with Layout A was recorded. Subsequently, with the aid of nanomanipulation, the nanowire was brought into contact with a nanopillar with a diameter of 395 nm. Finally, the nanowire was moved several micrometers away from the nanopillar and its PL emission was measured once again as a function of the pump power. The PL spectra recorded under illumination with an intensity of 234 kW/cm$^2$ for each configuration are plotted in Fig. 3a. A strong suppression of a longitudinal mode with a peak wavelength around 385.6 nm is observed in the emission spectrum when the nanowire is brought close to the nanopillar. The suppressed longitudinal mode reappears when the nanowire is moved away from the nanopillar. Fig. 3b shows SEM images of the measured nanowire and the nanopillar. Additional spectra and a comprehensive spectral analysis are shown in Supplement 1, Sec. 3.

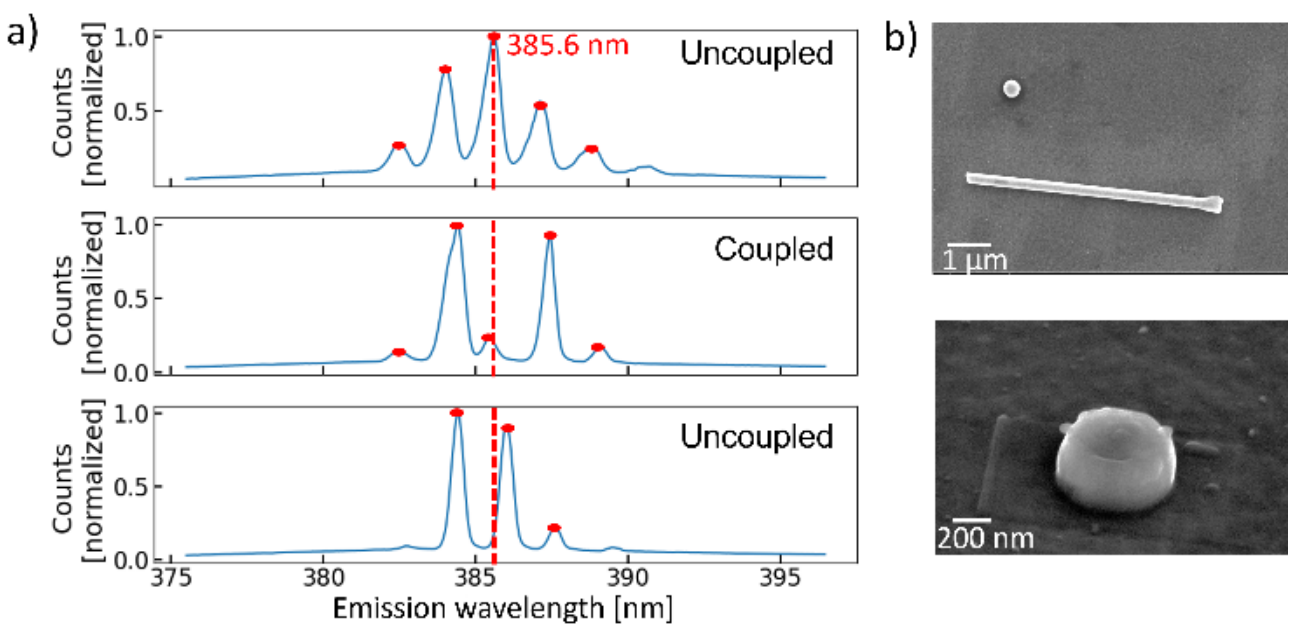

**Fig. 3** a) Sequence of PL spectra of the same nanowire (length of 6.1 µm, diameter of 255 nm, excited by a laser with an intensity of 234 kW/cm$^2$) in different positions with respect to a nanopillar (395 nm diameter). Upper: Nanowire placed by dry transfer on the sample away from the nanopillar. Middle: Nanowire brought into contact with nanopillar by nanomanipulator. Lower: Nanowire retracted from nanopillar. b) SEM pictures of the nanowire and the nanopillar.

We explain the mode suppression by performing FDTD simulations with the commercial software Lumerical [17], where the ZnO material is modeled as in Ref. [18] and $TiO_2$ was modeled with data from Ref. [15]. Fig. 4a shows the simulation setup: a $TE_{01}$ mode, which is believed to be the mode responsible for lasing at this diameter [6], is excited via a broadband mode source. A $TiO_2$ nanopillar is placed in contact with the nanowire and halfway between the nanowire end facets as the sketch in Fig. 4a shows. Transmission, scattering and absorption in the nanopillar are recorded and plotted in Figs. 4 b-d.
Figure 4b shows several modes emerging as transmission maxima at constant wavelengths for varying nanopillar diameter. They approximately reproduce the Fabry-Perot modes of the nanowire excited in the experiment. Their transmission is modulated as a function of nanopillar diameter. The nanowire material model has losses for wavelengths below 385 nm, thus artificially reducing transmission, scattering and absorption in the nanowire below this wavelength. For a pillar diameter of approximately 395 nm, as indicated by a dashed vertical white line, the transmission decreases for all excited longitudinal orders, including around a wavelength of 385.5 nm as indicated by a horizontal dashed white line. This decrease in transmission coincides with an increase in absorption (4c) while the scattered power decreases slightly (4d), pointing to coupling between the two systems. Thus, we conclude that the suppression of modes as observed in Fig. 3 is driven by resonant coupling of nanowire lasing modes to whispering-gallery modes in the nanopillar, which is signified by absorption maxima. The scattering reduces slightly because the confinement of radiation inside the nanopillar is stronger than in the nanowire.

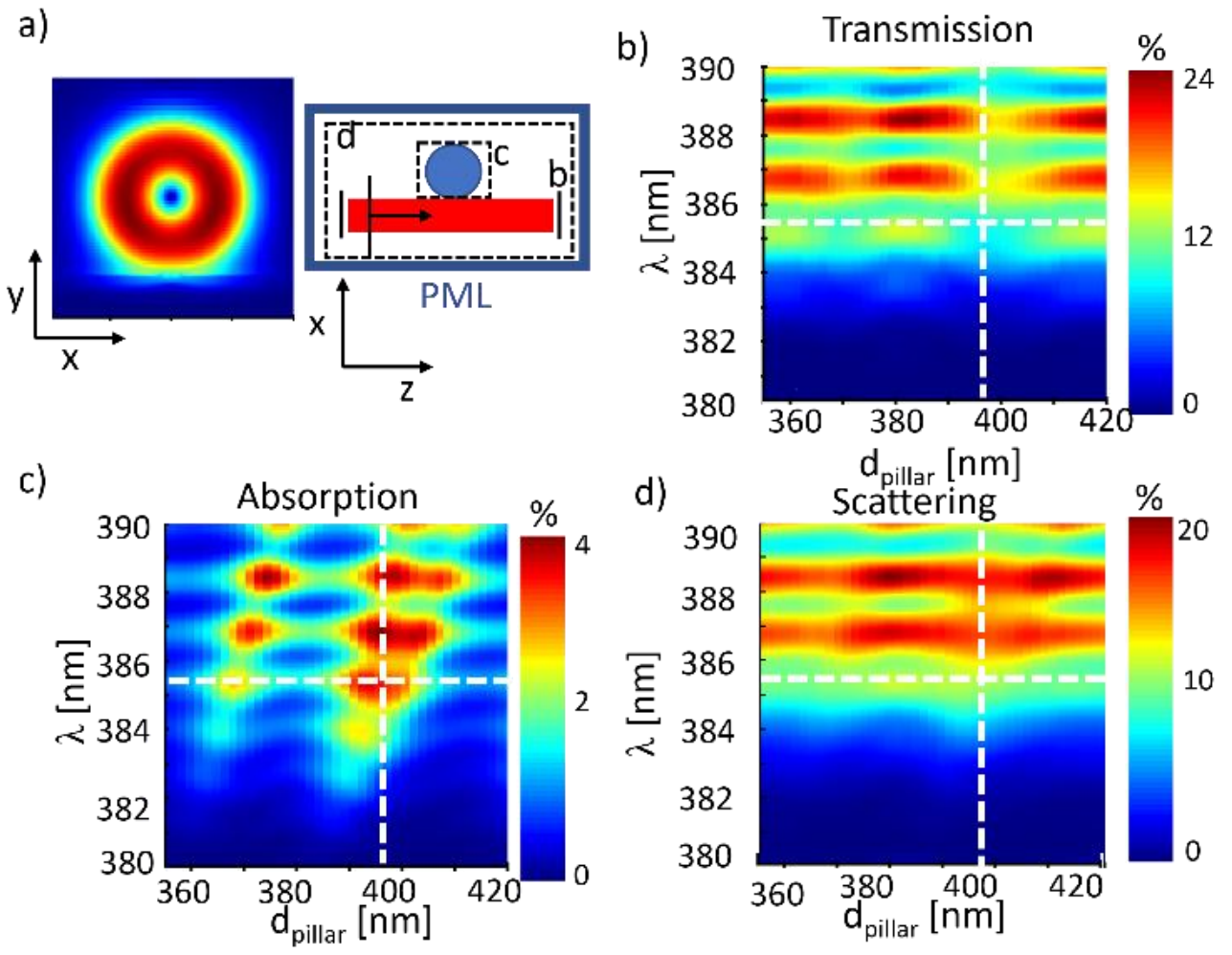

**Fig. 4** a) Sketch of the simulation configuration, showing the transversal profile of the excited mode and indicating the position of monitors used to extract b) transmission, c) absorption in the nanopillar and d) scattering for a nanowire of a diameter of 255 nm. The diameter of the nanopillar used in the experiment is indicated as a white dashed line in b, c and d. A dashed horizontal white line indicates the wavelength of the suppressed nanowire order.

Furthermore, we studied the interaction between a nanowire with a diameter of 160 nm and two nanopillars on a chip of Layout B. The results are plotted in Fig. 5a. The nanowire PL emission was first measured in a configuration where the nanowire was placed far away from any nanopillars. Subsequently the nanowire was brought in contact with a nanopillar (denominated as Nanopillar 1) having a diameter of 395 nm via nanomanipulation. The nanowire emission was recorded again and a suppression of longitudinal modes around a wavelength of 384-385 nm could be observed, matching the suppression window observed for the thicker nanowire. Then, the nanowire was moved again away from Nanopillar 1 and power-dependent PL spectra were recorded. The nanowire was finally moved to another nanopillar (denominated as Nanopillar 2, having a diameter of approximately 380 nm) and the nanowire's PL emission was measured, revealing mode suppression of several modes at lower wavelengths (383.3 nm and 381.5 nm). Additional spectra and a comprehensive spectral analysis are shown in Supplement 1, Sec. 3.

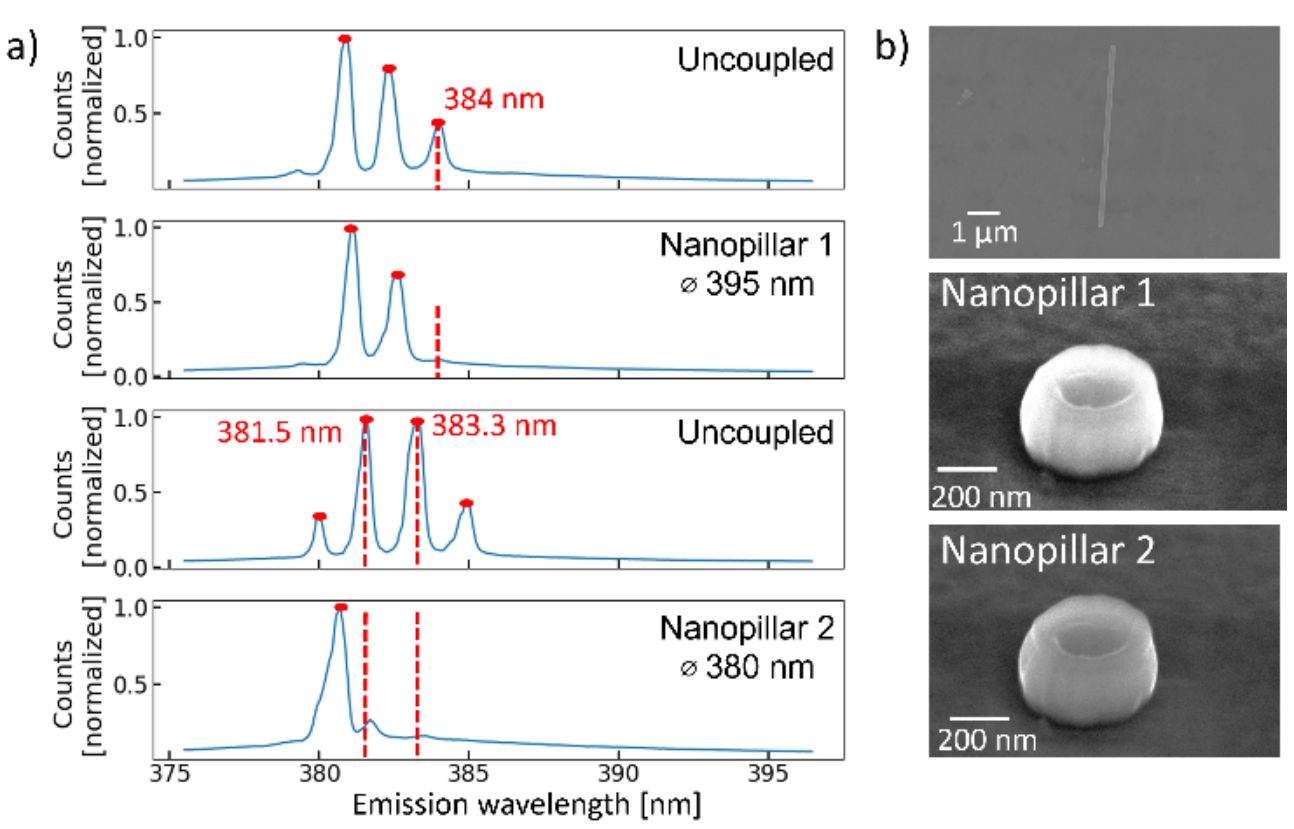

**Fig. 5** a) PL spectra of a nanowire with a length of 5.5 µm and a diameter of 160 nm excited by the pump laser at an intensity of 180 kW/cm² for different nanowire – nanopillar configurations. For the second and the last panel, the nanowire is brought into contact with nanopillars featuring an average diameter of 395 nm and 380 nm. b) SEM pictures of the nanowire and the nanopillars.

In order to explain these findings, FDTD calculations were performed and shown in Fig. 6. While the regime below a wavelength of 385 nm is not directly accessible by our simulations as a result of the passive material model, we can extrapolate the trends observed at higher wavelengths and compare with the measurements which show mode suppression at lower wavelengths. In the simulations, the nanowire diameter was set to 160 nm and the $HE_{11x}$ mode was chosen to excite the nanowire [6].

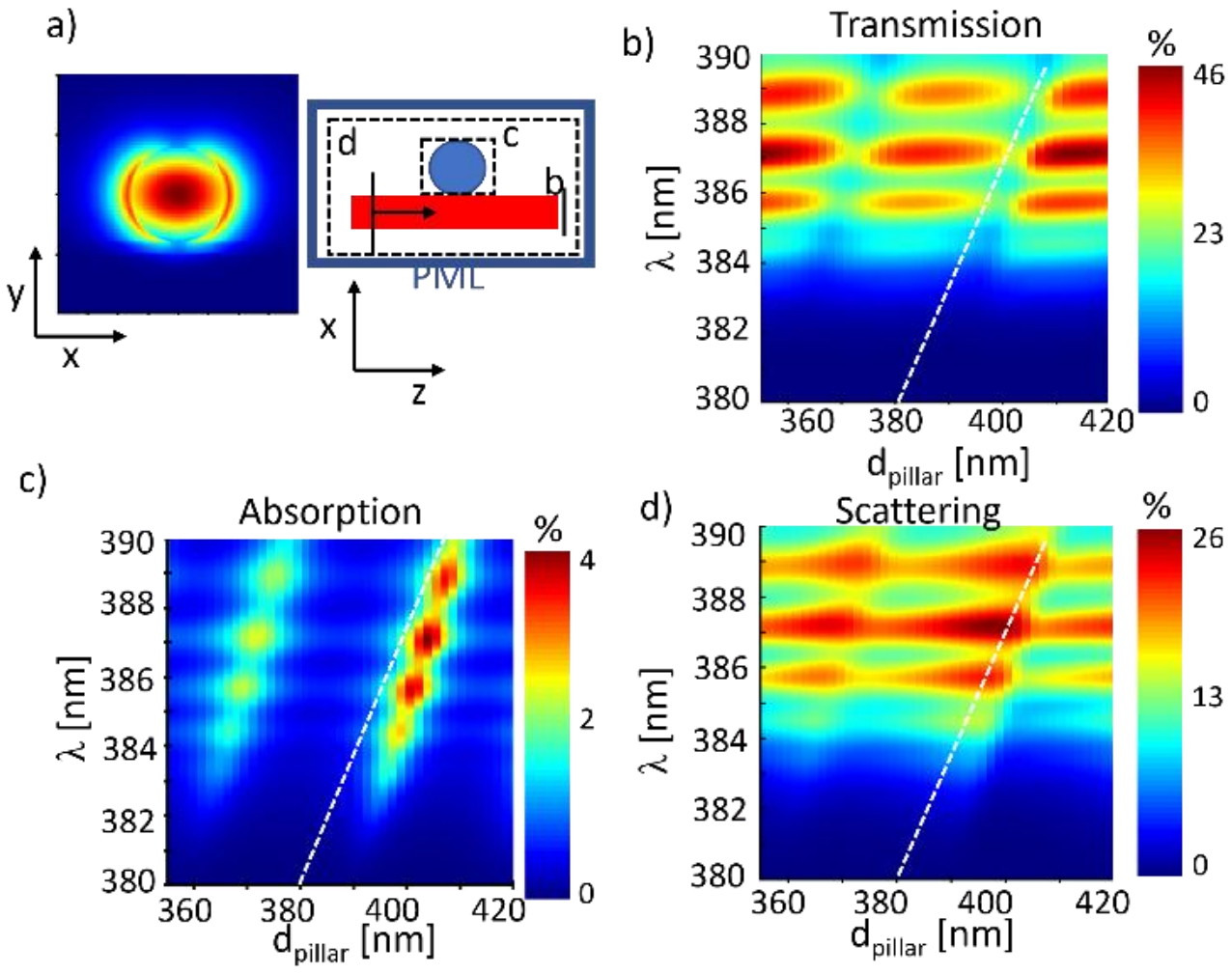

**Fig. 6** a) Sketch of the simulation configuration, showing the transversal field profile of the excited $HE_{11x}$ mode and indicating the position of the monitors used to extract b) scattering, c) nanopillar absorption and d) transmission of a nanowire of a diameter of 160 nm.

When extrapolating the feature showing reduced transmission in Fig. 6b to a wavelength of 384 nm, the resonant nanopillar diameter is about 395 nm. This corresponds to the diameter of Nanopillar 1 in the experiment. The observed effects (minimum of transmission, maximum of absorption and scattering) show a trend of decreasing resonance wavelength with decreasing nanopillar diameter. This is consistent with our experimental findings showing suppression of lower-wavelength Fabry-Perot modes for lower nanopillar diameters, albeit the passive material model is limiting quantitative comparisons.

In contrast to the observation presented in Fig. 4, the amount of scattered radiation does not decrease, but increases when the resonant nanopillar is chosen in Fig. 6. This is related to the fact that confinement of radiation in the mode excited in the nanopillar is weaker than in the nanowire.

At last, we calculate the electric field distribution in the lasing regime for both investigated nanowire diameters using Lumerical's built-in active material model on top of the passive ZnO model and a dipole excitation. The exact methodology is described in Ref. [6]. The field distributions are shown in Fig. 7.

The field distributions reveal that for the thick nanowire, the $TE_{01}$ mode is excited and coupled to an out-of-plane whispering gallery mode, whereas for the thinner nanowire, a $HE_{11x}$ mode is excited and coupled to an in-plane whispering gallery mode. This explains the differences in the magnitude of the radiation lost by scattering observed in the previous calculations: Whereas the $TE_{01}$ mode couples to a well-confined out-of-plane whispering-gallery mode, the $HE_{11x}$ mode, due to polarization matching, mainly couples to the in-plane whispering-gallery-mode.

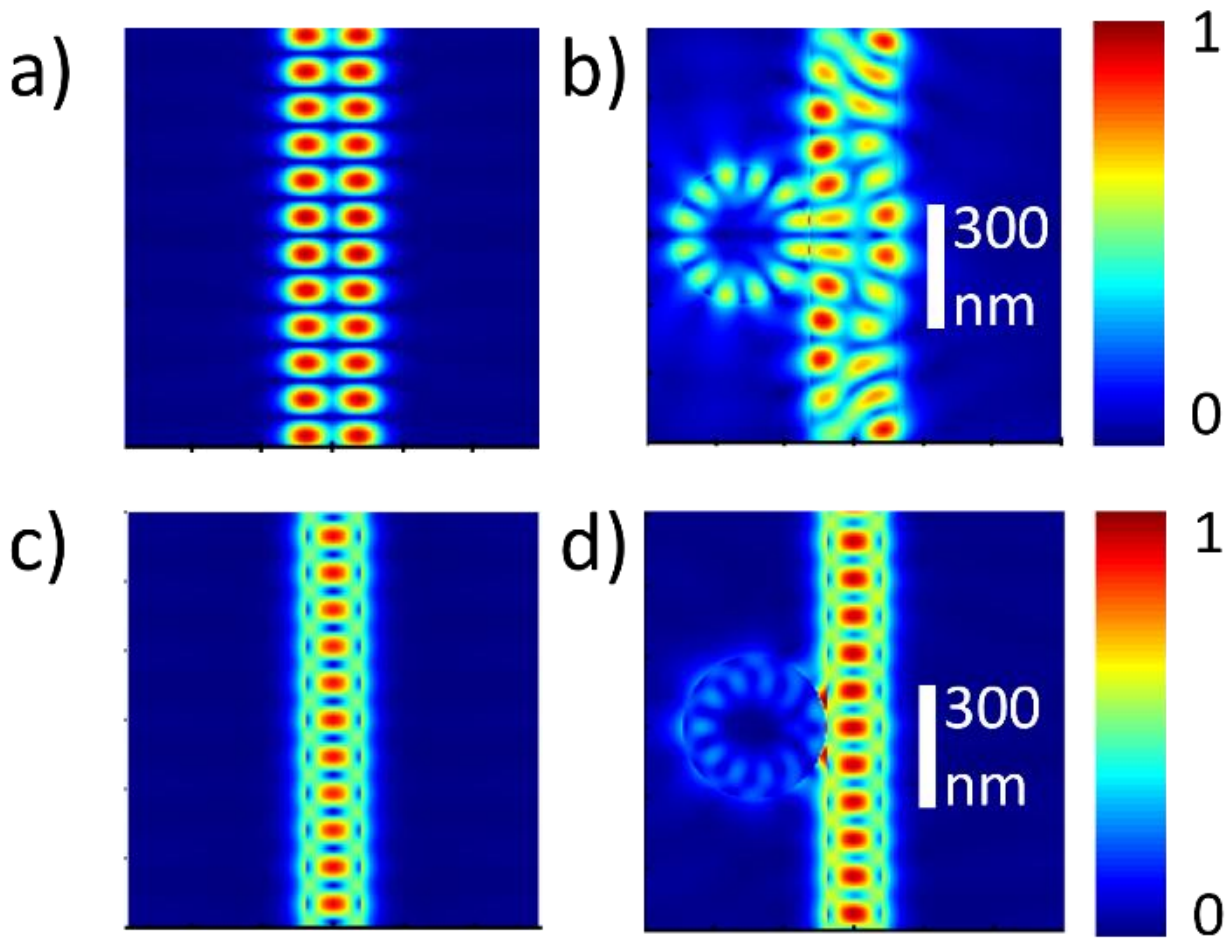

**Fig. 7** Electric field distribution in the lasing regime for a) an isolated nanowire with a diameter of 255 nm, b) a nanowire with a diameter of 255 nm next to a nanopillar with a diameter of 400 nm, c) an isolated nanowire with a diameter of 160 nm and d) a nanowire with a diameter of 160 nm coupled to a nanopillar with a diameter of 400 nm. The colorscales are identical to Figs. 4 and 6, scaled to the maximum values.

In this work, we demonstrated that selective suppression of longitudinal modes in ZnO nanowire lasers can be achieved via near-field coupling to $TiO_2$ nanopillars exhibiting resonances near the nanowire emission wavelength. Using FDTD simulations, we further confirmed that the suppression is caused by resonant interaction between the modes of the nanowire and the nanopillar. These findings offer a pathway toward mode engineering in nanowire-based photonic devices, particularly in applications requiring spectral purity or single-mode operation. The demonstrated approach may facilitate the integration of nanowire lasers into complex photonic circuits by enabling the tailored manipulation of their spectral emission properties. The experimentally achieved suppression effect is spectrally narrow, suggesting that further optimization of the platform are necessary for reaching the single-mode regime, for example by utilizing several nanopillars or broader resonances. Nevertheless, the demonstrated approach promises precise spectral control of nanowire lasing emission in future integrated devices.

**Back Matter**

**Funding.** This work was funded by the German Research Foundation (DFG) (CRC 1375 NOA "Nonlinear optics down to atomic scales"), project number 398816777 (subprojects B5, C1, C5 and Z3).

**Disclosures**. The authors declare no conflicts of interest.

**Acknoledgement**. The sample fabrication within this work was partially carried out by the microstructure technology team at IAP Jena. The authors would like to thank them for providing the fabrication facilities, carrying out processes and providing support.

**Data Availability Statement (DAS).** The raw data underlying this article is available upon request.

**Supplemental Document. See Supplement 1 for supporting content**.

## Full References

# Lasing Mode Control in ZnO Nanowires Coupled to $TiO_2$ Nanopillars: Supplementary 1

**Daniel Repp[1,*], Francesco Vitale[2], Raja hoffmann[2], Isabelle Staude[1,2], Thomas Siefke[1,3], Carsten Ronning[2], and Thomas Pertsch[1,3]**

[1]*Institute of Applied Physics, Abbe Center of Photonics, Friedrich Schiller University Jena, Albert-Einstein-Straße 15, 07745 Jena, Germany*
[2]*Institute of Solid State Physics, Friedrich Schiller University Jena, Max-Wien-Platz 1, 07743 Jena, Germany*
[3]*Fraunhofer Institute for Applied Optics and Precision Engineering, Albert Einstein Straße 7, 07745 Jena, Germany.*
**daniel.repp@uni-jena.de*

## 1. Sample fabrication

The samples were fabricated on single-side-polished Si(100) wafers (Siegert Wafer). Prior to deposition, the substrates were cleaned with Caro's acid in an Optiwet SB30 automated wafer-cleaning system. All subsequent layers were deposited without breaking vacuum using an Ionfab 300 LC ion-beam sputter deposition system (Oxford Instruments Plasma Technology). First, a 70 nm thick Al layer was deposited from an aluminum target using an ion energy of 1300 eV and an ion current of 250 mA. The substrate was rotated at 10 rpm, and the deposition time was 9 min 30 s. Before deposition, the substrate surface was pre-etched for 15 s at an ion energy of 400 eV. Subsequently, a 20 nm thick $Al_2O_3$ layer was reactively deposited from an aluminum target at an oxygen background pressure of $2\times10^{-4}$mbar. The deposition was performed at 1300 eV and 250 mA for 5 min 10 s while rotating the substrate at 10 rpm. Without breaking vacuum, a 200 nm thick $TiO_2$ layer was reactively deposited from a titanium target under the same ion-beam conditions for 128 min 10 s. Finally, an 80 nm thick Cr hard-mask layer was deposited from a chromium target for 20 min 20 s at 1300 eV and 250 mA, with the substrate rotating at 10 rpm. For electron-beam lithography, approximately 300 nm of FEP171 resist (Tokyo Ohka Kogyo, TOK) was applied using a UNIXX S700+ spin coater. The resist was exposed using a Vistec SB350OS electron-beam lithography system with an area dose of 9.5 µC $cm^{-2}$ and subsequently developed in OPD 4262. The resist pattern was transferred into the chromium layer by reactive-ion etching in an SI 591 system (SENTECH Instruments) using a chlorine–oxygen plasma. The remaining resist was subsequently removed in the same system using an oxygen plasma. The patterned chromium layer served as a hard mask for transferring the structures into the $TiO_2$ layer. This etching step was performed in an SI 500 C inductively coupled plasma etching system (SENTECH Instruments) using a $CHF_3/SF_6/O_2/Ar$ plasma at an ICP power of 700 W. After pattern transfer, the remaining chromium hard mask was removed in the SI 591 system using a chlorine–oxygen plasma. Finally, the wafer was diced into individual samples.

## 2. Micro-photoluminescence (µ-PL) Setup

The room-temperature optical characterization of individual nanowires in the single-nanowire scheme was accomplished by optically exciting the nanowires by slightly defocusing the radiation emitted by a frequency-tripled Nd:YAG laser ($\lambda_{exc}$ = 355 nm, $f_{rep}$ = 100 Hz, $\tau_{pulse}$ = 7 ns) through a 100x NUV objective (NA = 0.53) down to a spot size of ≈ 10 µm, so to illuminate the whole nanowire. The acquisition of the photoluminescence (PL) spectra was carried out by using a spectrometer (Princeton Instruments SP-2500i) equipped with a 1200 lines/mm grating (blazed at 300 nm), and connected to a nitrogen-cooled, front-illuminated CCD camera. The PL measurements were all accomplished in the same configuration, namely with the pump polarization kept parallel to the nanowire *c*-axis. The single-nanowire μ-PL measurements were aided by a nanomanipulation technique enabled by a remotely-controlled nanomanipulator (Kleindiek MM3A-EM) featuring a tungsten (W) tip with a radius of curvature of $r \approx 1$ μm, installed in the μ-PL setup between the sample and the objective. The nanowire manipulation was performed in air while being monitored under an optical microscope instead of an electron microscope to avoid any modification of the optical properties (waveguiding, lasing threshold, etc.) induced by a prolonged exposure to an electron beam [1].

### 3. Additional analysis of lasing spectra

In the main manuscript, the spectra are only plotted for one value of the excitation power density for clarity. Here, we present the spectra collected for different excitation power densities as well as an analysis of the spectral linewidth and the total emission as a function of excitation power, which reveals the lasing threshold. These spectra are plotted in Fig. S1 and the spectral analysis in Fig. S2 for the first set of experiments featuring the nanowire with a diameter of 255 nm.

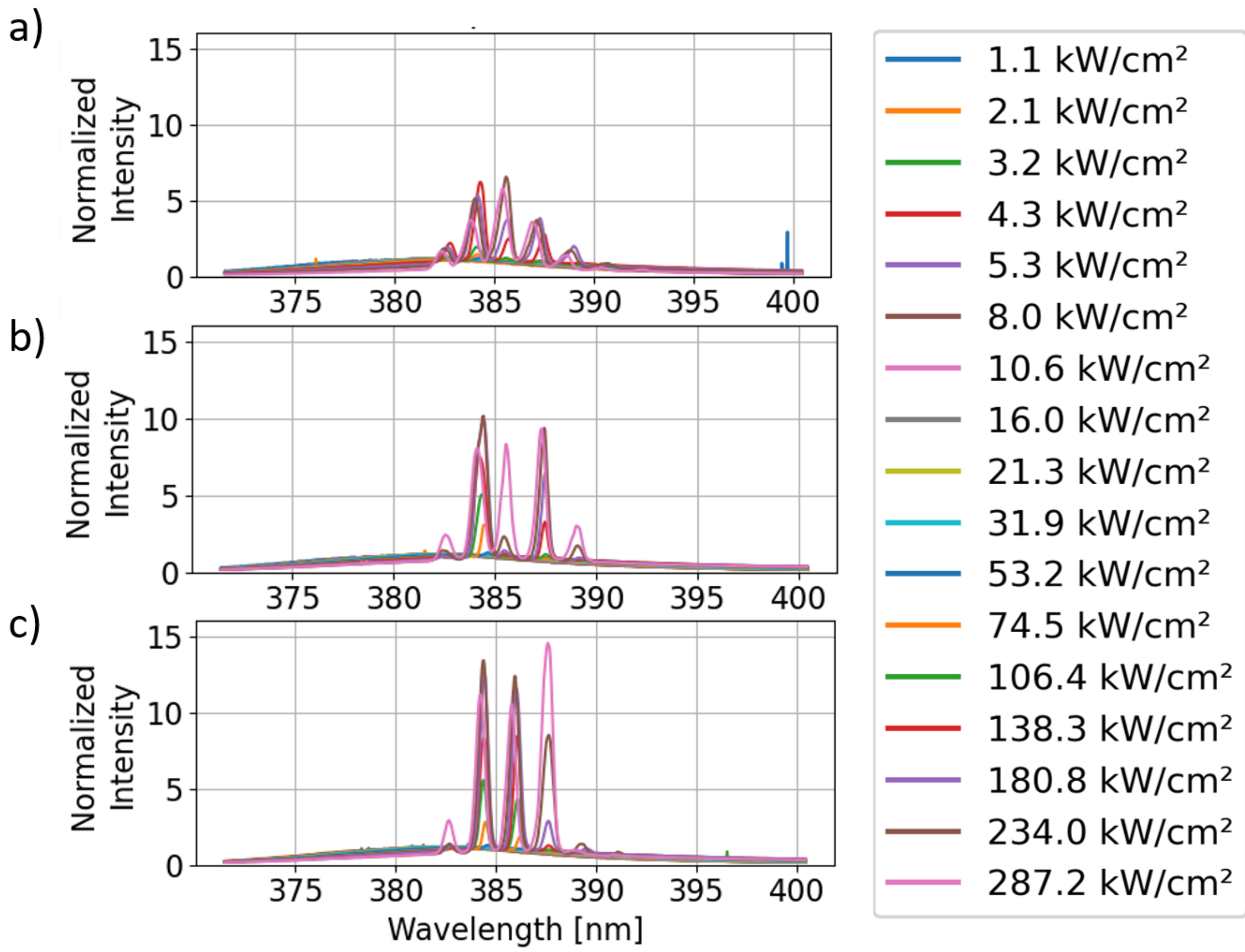

Fig. S1: Intensity-dependent emission spectra for a nanowire with a diameter of 255 nm uncoupled (a and c) and coupled (b) to a nanopillar with a diameter of 395 nm. The spectra are normalized to the maximum of the background gaussian fit.

The spectra shown in Fig. S1 reveal that the mode-suppression effect persists for a large range of excitation power densities. It can be overcome by an increase of excitation density, proving that the suppression mechanism is a result of increased lasing threshold for individual Fabry-Perot modes of the nanowire.

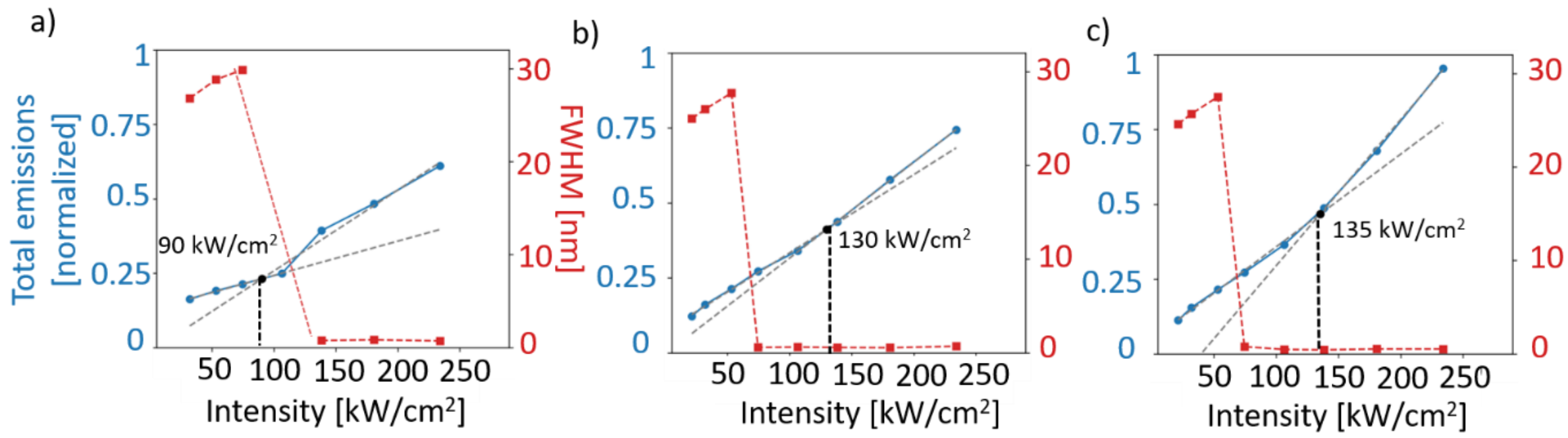

Fig. S2: Total emission and spectral FWHM as a function of excitation intensity for a nanowire with a diameter of 255 nm uncoupled (a and c) and coupled (b) to a nanopillar with an average diameter of 395 nm. The lasing threshold is determined as the crossing between two linear fits to the spontaneous emission and the stimulated emission regime, marked in the plots.

The spectral analysis reveals that the lasing threshold increases appreciably when coupling the nanowire to the nanopillar. However, when moving of the nanowire away from the nanopillar, the threshold maintains its elevated value. Therefore, we assume that the increase in lasing threshold is related to the nanomanipulation and not to a change in the mode properties of the nanowire laser.

We plot furthermore in Fig. S3 the spectra and in Fig. S4 the emission and spectral linewidth for the nanowire with a diameter of 160 nm.

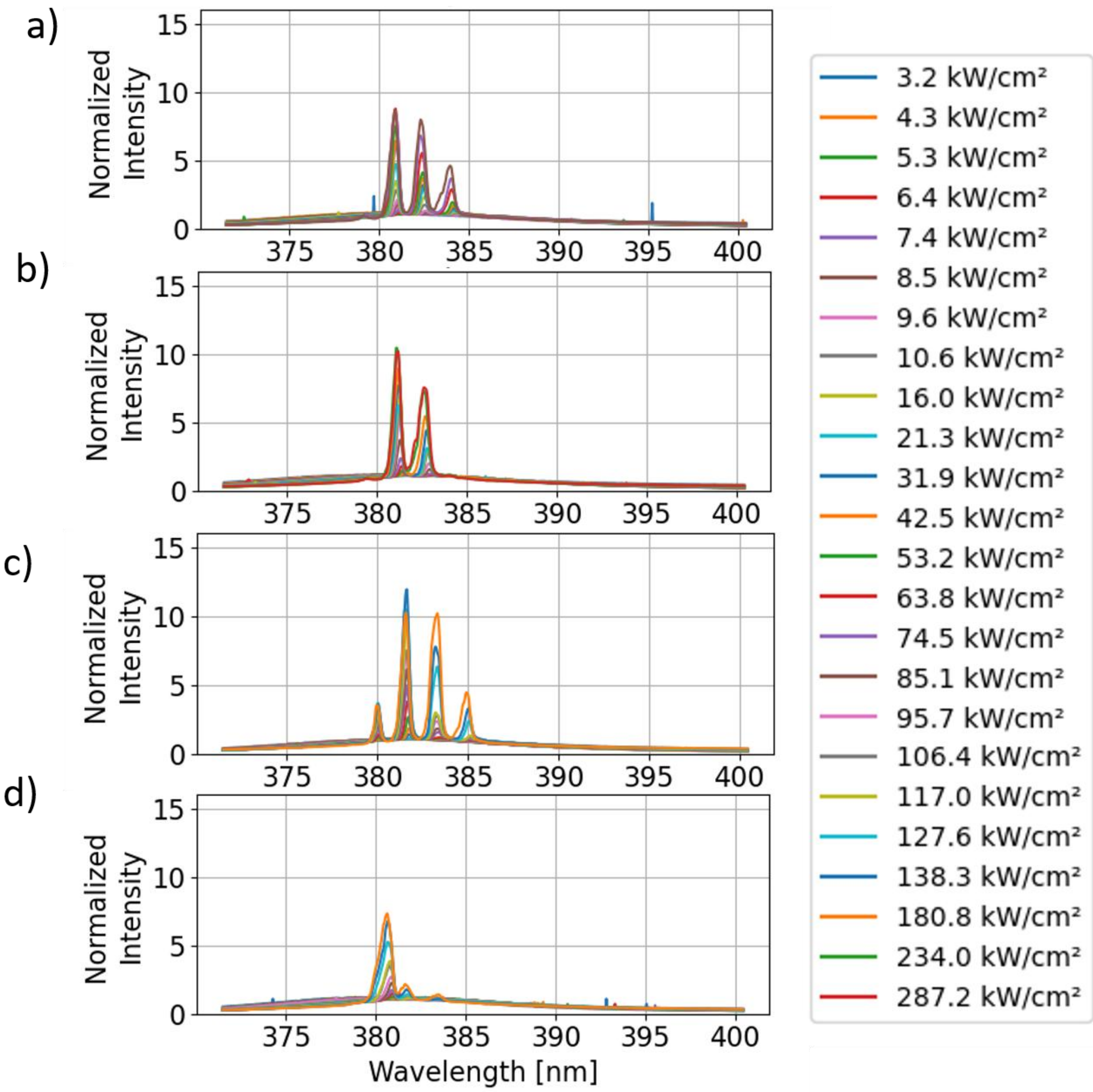

Fig. S3: Emission spectra depending on the excitation intensity for a nanowire with a diameter of 160 nm uncoupled (a and c) and coupled to a nanopillar with an average diameter of b) 395 nm or d) 380 nm. The spectra are normalized to the maximum of the background gaussian fit.

The emission spectra shown in Fig. S3 reveal the same suppression effect shown in the main manuscript. The main difference to Fig. S1 is that this effect is observed for all excitation intensities. This points towards a stronger suppression to be achievable in thin nanowires as a result of the stronger scattering channel.

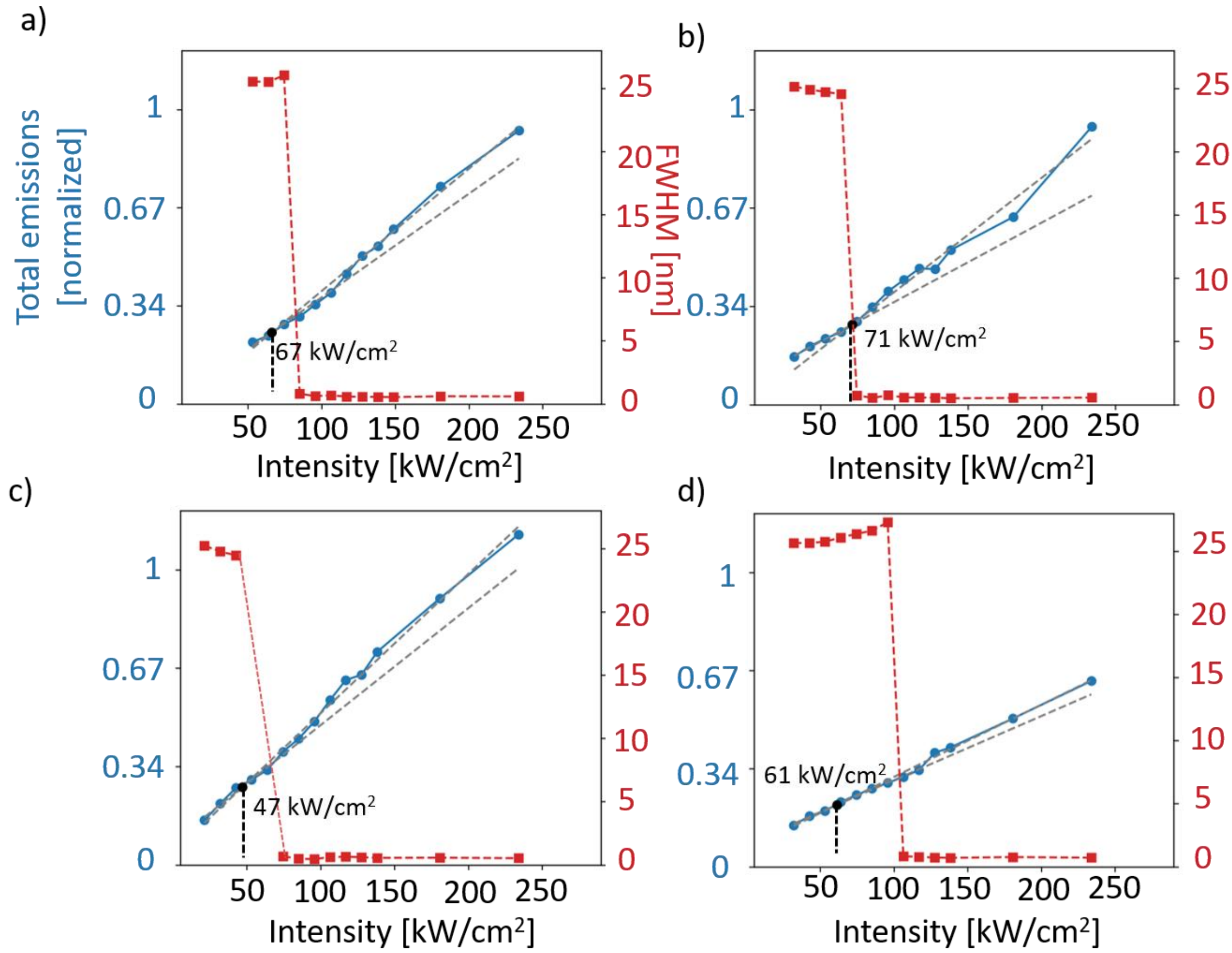

Fig. S4: Total emission and spectral FWHM as a function of excitation intensity for a nanowire with a diameter of 160 nm uncoupled (a and c) and coupled to a nanopillar with an average diameter of b) 395 nm or d) 380 nm. The lasing threshold is determined as the crossing between two linear fits to the spontaneous emission and the stimulated emission regime, marked in the plots.

The spectral analysis in Fig. S4 reveals no significant changes in lasing thresholds in the different configurations.